\documentclass[twocolumn,twoside]{revtex4}
\usepackage{graphicx}
\usepackage{fancyhdr}
\usepackage{siunitx}

\begin{document}

\title{FCC-ee, an Accelerator for New Physics Searches in the Heavy Quark Sector}

%

\author{R.~Novotný}
\affiliation{Department of Physics and Astronomy, University of New Mexico, Albuquerque, NM, USA, 87131}

\begin{abstract}
This manuscript is devoted to the description of the proposed Future Circular Collider (FCC) project and its physics program focused on measurements involving heavy quarks.
It summarizes the report submitted to the US 2021 Snowmass Process and European Strategy on Particle Physics.
The proposed electron-positron collider based on established technologies will provide high instantaneous luminosities at center-of-mass energies from the Z resonance through the ZH and WW and up to the $t\bar{t}$ threshold. 
This will allow a very rich set of fundamental measurements as well as the study of heavy-flavor and tau physics in ultra-rare decays beyond the LHC reach.

\end{abstract}

\maketitle

\thispagestyle{fancy}


\section{Introduction}
The near future of heavy-flavor physics research will be covered by the LHC \cite{Evans:2008zzb} and Belle~II \cite{Belle-II:2018jsg} experiments. The LHC started Run~3 operation in 2022 that will take 3 years, and after that, the HL-LHC upgrade will provide \SI{3000}{\per\femto\barn} of $pp$ collisions at \SI{14}{\TeV}. The Belle~II experiment at SuperKEKB started to take data in 2019; they will collect high statistics on the b-hadrons from $\Upsilon(4S)$ decays, and they aim to collect about \SI{50}{\per\atto\barn} over the six-year lifetime.
It is hoped that these two experiments will make a great contribution to the precise measurements of weak interaction parameters and find New Physics (NP) beyond the Standard Model (SM) of particle physics. 
However, they are not able to cover all possible measurements, especially in some decays involving heavier particles like the $B_s$, $B_c$ and $\Lambda_b$ that are beyond the reach of the Belle~II experiment and might be very complicated at the LHC due to high contamination. 
There is a large effort to design and build new Higgs factories like the ILC, CLIC, CEPC, and FCC, and the B-physics research program can benefit from this effort. 
This paper is focused on the Future Circular Collider (FCC) project and summarizes the report submitted to the US 2021 Snowmass Process \cite{Bernardi:2022hny} and European Strategy on Particle Physics \cite{Benedikt:2653673}.
In the following sections, the FCC project will be explained together with the part of its physics program that is focused on measurements involving heavy quarks.

\section{FCC accelerator and instrumentation}
The critical infrastructure of the FCC project is a tunnel $>\SI{90}{\km}$ in circumference that will be located close to the CERN laboratory near Geneva. 
The main goal is to build an electron-positron collider (denoted as FCC-ee) with the possibility to build a hadron collider (called FCC-hh) after the physics program of the FCC-ee has been completed.
The FCC-ee collider design is now being developed for either 2 or 4 symmetric interaction points located at four of the access points and with radiofrequency (RF) cavities, collimation, injection, and extraction occupying the other 4 straight sections.
The design is robust enough to provide high luminosity over the center-of-mass energy range from 90 to \SI{365}{\GeV}.
The FCC-ee collider has separate rings for electrons and positrons, and the design is optimized for \SI{50}{\MW} synchrotron radiation energy loss per collider ring across the operating energy range.
The baseline configuration is based on \SI{400}{\MHz} RF systems with Nb/Cu cavities; however,
to reach operation at the $t\bar{t}$ threshold, the system needs to be augmented by an \SI{800}{\MHz} RF system with bulk-Nb cavities.
The operation model of the FCC-ee collider starts with a \SI{91}{\GeV} run to study the Z boson with very high luminosity and than incrementally increases the energy every couple of years to reach \SI{160}{\GeV} for WW, \SI{240}{\GeV} for ZH, and \SI{356}{\GeV} for $t\bar{t}$ production. More details can be found at conceptual design reports \cite{FCC:2018evy,FCC:2018vvp}.

\begin{figure}[h]
\centering
\includegraphics[width=\linewidth]{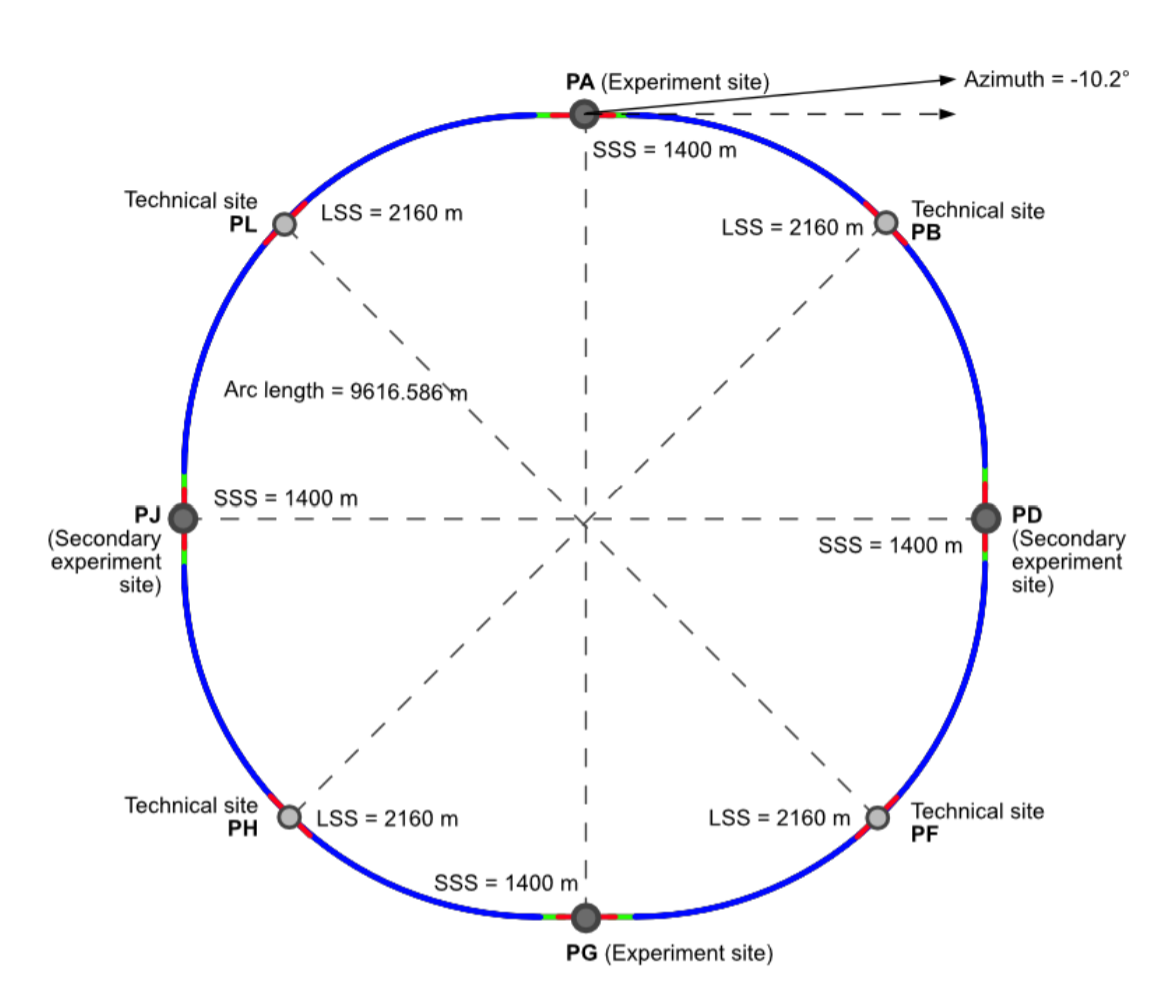}
\caption{The FCC layout referred to as PA31-1.0; four possible experiments could
be located at PA, PD, PG, and PJ while RF stations would be located PH and PL and
injection/extraction and collimation could be located in the PB and PF straights. Taken from Ref. \cite{Bernardi:2022hny}.} \label{FCC_layout}
\end{figure}

To carry out an extensive physics program, high granularity detectors need to be built. 
Two complementary detector design concepts have been proposed for FCC-ee, the ``CLIC-like Detector" (CLD) \cite{Bacchetta:2019fmz} and the ``International Detector for Electron-positron Accelerator" (IDEA) \cite{Bedeschi:2021nln}. 
The concepts are evolutions of the detectors for the past and current colliders incorporating the latest results from years of R\&D as well as the newest technologies.
Both detector concepts feature a \SI{2}{\tesla} solenoidal magnetic field.
The CLD detector features a silicon pixel vertex detector, a silicon tracker, followed by a highly granular calorimeter (a silicon-tungsten ECAL and a scintillator-steel HCAL) surrounded by a superconducting solenoid and muon chambers interleaved with steel return yokes.
The IDEA detector comprises a silicon vertex detector, a large volume extremely light drift chamber surrounded by a layer of silicon detectors, a thin
low-mass superconducting solenoid, a pre-shower detector, a dual-readout fiber calorimeter, and muon chambers within the magnet return yoke.

\section{Flavor physics opportunities at the FCC}
The Z run of the FCC-ee will provide unprecedentedly high statistics of $\mathcal{O}(5\times10^{12})$ Z bosons decaying through $Z\rightarrow \bar{b}b$ and $Z\rightarrow \bar{c}c$ events that will be recorded without any triggers or prescales. 
This gives an opportunity to enrich the knowledge of flavor physics of quarks and leptons.
The flavor program of the FCC-ee experiment will be a natural continuation of the upgraded LHCb experiment \cite{LHCb:2018roe} and the Belle~II experiment \cite{Belle-II:2018jsg}.
The anticipated production yields of heavy-flavored particles at FCC-ee are compared to the Belle~II experiment in Table~\ref{tab:events}.

\begin{table}[!ht]
\caption{Expected production yields of heavy-flavored particles at Belle~II (\SI{50}{\per\atto\barn}) and FCC-ee (Z pole). The $X/\bar{X}$ represents the production of a B-hadron or its charge conjugated state. The Z branching fractions and hadronization rates are taken from Ref. \cite{ParticleDataGroup:2020ssz}.}
\begin{tabular}{| c | c | c |}
\hline
Particle production ($10^9$)& Belle~II& FCC-ee \\\hline
$B^0/\bar{B}^0$& 27.5 & 620 \\
$B^+/B^-$& 27.5 & 620 \\
$B_s^0/\bar{B}_s^0$& n/a & 150 \\
$B_c^0/\bar{B}_c^0$& n/a & 4\\
$\Lambda_b/\bar{\Lambda}_b$ & n/a & 130 \\
$c/\bar{c}$& 65 & 600\\
$\tau^+/\tau^-$& 45 & 170\\\hline
\end{tabular}

\label{tab:events}
\end{table}

The possibilities of flavor measurement are not restricted to a Z pole run only.
The decays of on-shell W bosons will provide a particularly rich laboratory for
studies of the CKM matrix, and the HZ run will provide valuable data for the charged-lepton flavor violating (cLFV) decays, as described in more detail below.

\subsection{Flavor-changing neutral currents}
In the SM of particle physics, the electroweak couplings of leptons to gauge bosons are independent of their flavor; the model is referred to as exhibiting lepton universality (LU). 
Flavor-changing neutral-current (FCNC) processes, in which a quark changes its flavor without altering its electric charge, provide an ideal laboratory to test LU.
The SM forbids FCNCs at tree level and only allows amplitudes involving electroweak loop Feynman diagrams. 
The absence of a dominant tree-level SM contribution implies that such transitions are rare, and therefore sensitive to the existence of new particles.

The most promising measurements of FCNC in recent years are connected with the semileptonic decays $B \to K^{(*)}e^+e^- $and $B \to K^{(*)}\mu^+\mu^- $. 
These measurements show a number of persistent $2\sigma - 3\sigma$ tensions between data and SM expectations, in particular in the lepton flavor universality ratios $R_{K^{(*)}}$ \cite{LHCb:2017avl, LHCb:2019hip, LHCb:2021trn} and the angular observable $P_5^\prime$ \cite{LHCb:2020lmf}.
The FCC-ee measurement can provide an independent confirmation of these so-called B ``anomalies'' and if confirmed, the anomalies in the rare B decays establish a generic new physics scale of $\sim \SI{35}{\TeV}$.

\begin{figure}[h]
\centering
\includegraphics[width=\linewidth]{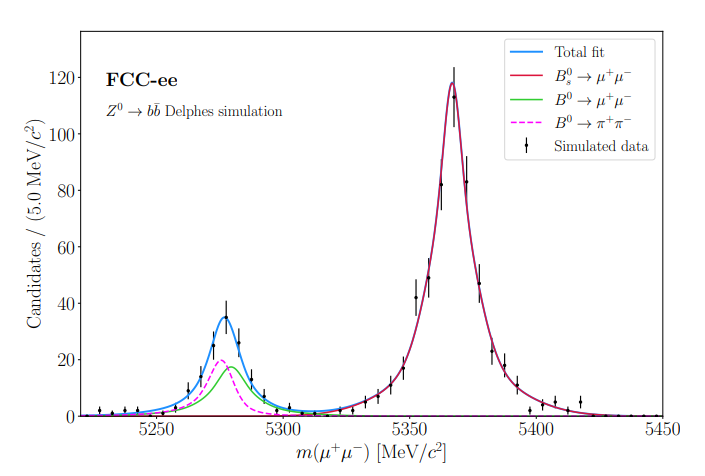}
\caption{ Reconstructed invariant mass of $B^0 \rightarrow\mu^+\mu^-$ and $B^0_s \rightarrow\mu^+\mu^-$ signals at the FCC-ee based on $10^{13}$ Z decays. Taken from Ref. \cite{Monteil:2021ith}.} \label{BllFCC}
\end{figure}

The other measurement of FCNC that merits attention is in the leptonic decays $B_s\rightarrow\mu^+\mu^-$ and $B^0 \rightarrow\mu^+\mu^-$. These have extremely small branching ratios in the SM of $(3.66\pm0.14)\times10^{-9}$ and $(1.03\pm0.05)\times10^{-10}$, respectively \cite{Beneke:2019slt}.
Their precise but tiny branching ratios make them highly sensitive probes of new physics.
Observation of $B^0 \rightarrow\mu^+\mu^-$ is within reach of the HL-LHC, but the advantage of FCC-ee over the LHC is the excellent mass resolution that allows a clear separation of the $B_s$ and $B^0$ signals in the di-muon invariant mass spectrum, as is shown in Figure~\ref{BllFCC}.
It is expected that with FCC-ee Z pole run statistics, $\sim 540$ $B_s \rightarrow \mu^+\mu^-$ events and $\sim 70$ $B_d \rightarrow \mu^+\mu^-$ events will be reconstructed in the SM scenario \cite{Monteil:2021ith}.

\begin{figure}[h]
\centering
\includegraphics[width=\linewidth]{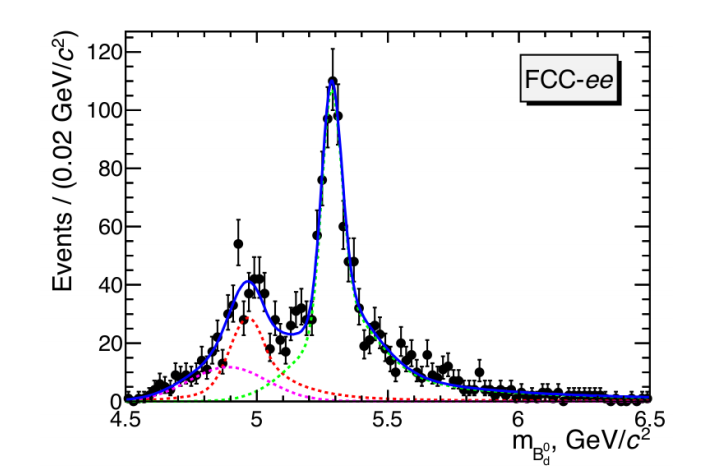}
\caption{Invariant mass reconstruction of $\bar{B}^0 \to K^{*0}\tau^+\tau^- $ candidates at the FCC-ee based on $10^{13}$ Z decays. Taken from Ref. \cite{Kamenik:2017ghi}.} \label{BKllFCC}
\end{figure}

Both previous measurements can benefit from FCC-ee unique sensitivity to tauonic and semitauonic decays. 
Current bounds on the branching ratios of decays like, $B_s\rightarrow\tau^+\tau^-$, $B_0\rightarrow\tau^+\tau^-$, and $B \rightarrow K^{(*)}\tau^+\tau^-$, are still several orders of magnitude above the SM predictions.
Sensitivities will improve at the HL-LHC and Belle~II but cannot reach the SM precision.
The $B \rightarrow K^{(*)}\tau^+\tau^-$ decay is a particularly rich laboratory to
probe new physics since many BSM scenarios predict characteristic effects in the decays with taus in the final state. 
Using three-prong tau decays and assuming high precision vertex reconstruction, it has been shown that this decay can be fully reconstructed at Z-pole machines, and about $\mathcal{O}(10^3)$ cleanly reconstructed SM events are expected at the FCC-ee as is shown in Figure~\ref{BKllFCC}.
This sample size will not only allow a precision measurement of the
$B \rightarrow K^{(*)}\tau^+\tau^-$ branching ratio, but also opens up the possibility of measuring the angular distribution of the decay.

Another group of analyses can be performed in the channels with missing energy. 
The FCNC decays $B \rightarrow K\nu\bar{\nu}$ and $B \rightarrow K^{*}\nu\bar{\nu}$ are well established probes of new physics.
Belle~II is expected to make the first observation of these decays and measure their branching ratios with an uncertainty of $\sim 10\%$.
FCC-ee should be able to further improve these measurements, which are highly motivated given that they are theoretically well understood.
FCC-ee has the unique opportunity to measure the related decays $B_s \rightarrow \phi\nu\bar{\nu}$, $\Lambda_b \rightarrow \Lambda\nu\bar{\nu}$, and even $B_c \rightarrow D_s\nu\bar{\nu}$. 
Combining the results from the whole family of $b \rightarrow s\nu\bar{\nu}$ decays will be a powerful way to probe BSM physics.

The FCNF studies in the decays involving the $B_c$ meson are still largely uncharted territory.
The $B_c$ mesons are not produced at Belle~II and are challenging to reconstruct at hadron colliders.
It is expected to get about $10^9$ $B_c$ mesons from the Z-pole run of the FCC-ee, which puts the FCC-ee in a unique position to make precision measurements of $B_c$ decays.
Very interesting are the theoretically clean leptonic decays $B_c \rightarrow \tau\nu$ and $B_c \rightarrow \mu\nu$ that have new physics sensitivity. 
This complements the well studied decay modes $B \rightarrow \tau\nu$ and $B \rightarrow D^{(*)}\tau\nu$.
The ratio $\mathcal{B}(B_c \rightarrow \mu\nu)/\mathcal{B}(B_c \rightarrow \tau\nu)$ is of particular interest in view of the existing anomalies in the lepton flavor universality ratios $R_D^{(*)}$.
Limited final state information renders a measurement of $B_c \rightarrow \tau\nu$ infeasible at hadron colliders.
Using three-prong tau decays, the study finds that a high purity signal sample
containing $\sim4000$ $B_c \to \tau\nu$ decays is achievable.
The channel $B_c \to \mu\nu$ is suppressed by a factor $\sim m^2_\mu/m^2_\tau \sim 3 \times 10^{-3}$,
but we can still expect $\mathcal{O}(10^5)$ $B_c \to \mu\nu$ events at the FCC-ee. Precision measurement would require good control over the background decays.

\subsection{Precise CKM and $C\!P$-violation parameter studies}
The CKM matrix induces flavor-changing transitions inside and between generations in the charged sector at the tree level.
Many of the observables in $C\!P$-violation studies are very precisely predicted, so they warrant continued experimental attention.
All current experimental results exhibit remarkable agreement with the SM predictions; however, they still leave room for BSM contributions to $C\!P$-violating transitions. 
The measurements, which become limited by systematic biases at the LHC, will benefit from the cleaner and very different environment of the FCC-ee, especially decay modes involving $B_s$, $B_c$, or b-baryons with neutral final state particles.
Due to better particle identification, it is expected that flavor-tagging efficiency will be significantly higher than in the LHC era. This will be a large advantage for any time-dependent measurement.

A particular strength of the FCC-ee flavor program will be the ability to make very sensitive studies of decays containing neutrals. 
This possibility will enable measurement of various $C\!P$-violating asymmetries such as the time-dependent $C\!P$ asymmetry in the $B^0 \rightarrow \pi^0 \pi^0$ decay and measurement of $C\!P$ asymmetry in $B^-\rightarrow DK^-$ (where D indicates a superposition of $D^0$ and $\bar{D}^0$) and $B_s \rightarrow D_s^{(*)\pm} K^\mp$. 
Another benefit of the FCC-ee environment will be the possibility to measure semileptonic $C\!P$-violating asymmetries and determine $|V_{ub}/V_{cb}|$ with $B_s^0$ mesons and $\Lambda_b$ baryons that are not accessible at current experiments with enough precision.

The measurement of $C\!P$ violation parameters in $B_\mathrm{d,s}$ meson mixing has been well established in many b hadron decays.
These systems are very sensitive to any new BSM contribution because the box diagrams that drive the oscillations and carry $C\!P$-violating phases are the neutral entry point for any new BSM particle. 
To probe deviations from the SM predictions and test the BSM models, increased precision
on $|V_{ub}|$ and $\gamma$ is required \cite{Charles:2015gya}.
For the sample size to be collected at the FCC-ee, it will be possible to measure the relevant observables with precision similar or better than that of previous experiments 
\cite{FCC:2018byv, Barbieri:2021wrc}. 
The measurement of the very small asymmetries $a_{sl}=\frac{\Gamma(\bar{B}_q^0\to\bar{f})-\Gamma(B_q^0\to f)}{\Gamma(\bar{B}_q^0\to\bar{f})+\Gamma(B_q^0\to f)}$ is very valuable in providing sensitivity to $h_d$ and $h_s$, due to the precision of the prediction.
It is expected that the first observation of $C\!P$-violation in B mixing will be within reach and the analysis of the BSM contributions in box mixing processes will provide a valuable test of BSM physics. 
Assuming the minimum flavor violation scenario, where the new flavor structures are aligned with the SM Yukawa couplings, the energy scale up to \SI{20}{\TeV} can be probed.

\begin{figure}[!th]
\centering
\includegraphics[width=\linewidth]{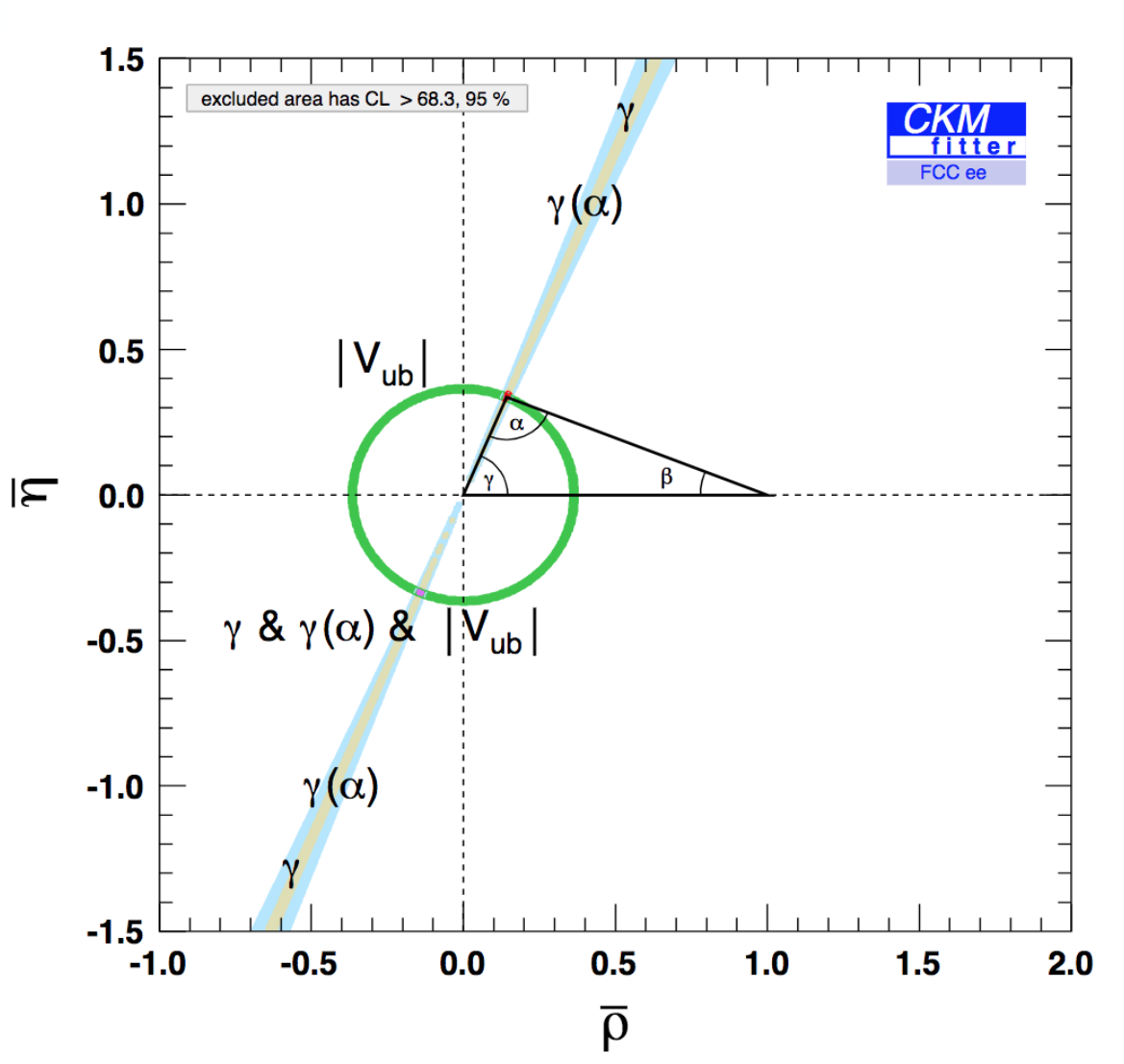}
\caption{Possible status of the Unitarity Triangle constraints in the ($\bar{\rho} ,\bar{\eta}$) plane after a FCC-ee Z-pole run. The red hatched region of the global combination corresponds to 68\% CL. Taken from Ref. \cite{Charles:2004jd}.} \label{CKM}
\end{figure}

\subsection{Charged-lepton flavor violating decays}
The charged-lepton flavor violating (cLFV) decays are transitions among $e$, $\mu$, and $\tau$ that do not conserve the lepton family number.
Evidence of cLFV would be a clear signal of new physics and would directly address the physics of flavor and generations.
The B-meson decay channels in which the flavor anomalies are observed are polluted by complicated strong dynamics, while the cLFV decays are much cleaner.
This suggest, that the cLFV modes will provide a better means to study the mechanism generating lepton flavor violation or non-universality once they are discovered.

The Higgs boson flavor violating decays to leptons provide a perfect system for cLFV studies.
The present best direct limits on the branching fractions of $H \to e\mu$, $H \to e\tau$, and $H \to \mu\tau$ decays are $6\times 10^{-5}$, $2.2\times10^{-3}$, and $1.5\times10^{-3}$ at $95\%$ CL, respectively \cite{ATLAS:2019old, CMS:2021rsq}.
With about one million Higgs bosons produced in association with the Z boson at the FCC-ee, about the same sensitivity in the $e\mu$ channel and about a factor of two better sensitivity in the other two channels relative to full HL-LHC running can be obtained.
Using all decay channels of the $\tau$ lepton, in particular hadronic decay modes and the cleaner decay modes of the Z boson, may result in further improvement of the FCC-ee sensitivity to these LFV processes.

\begin{figure}[!th]
\centering
\includegraphics[width=\linewidth]{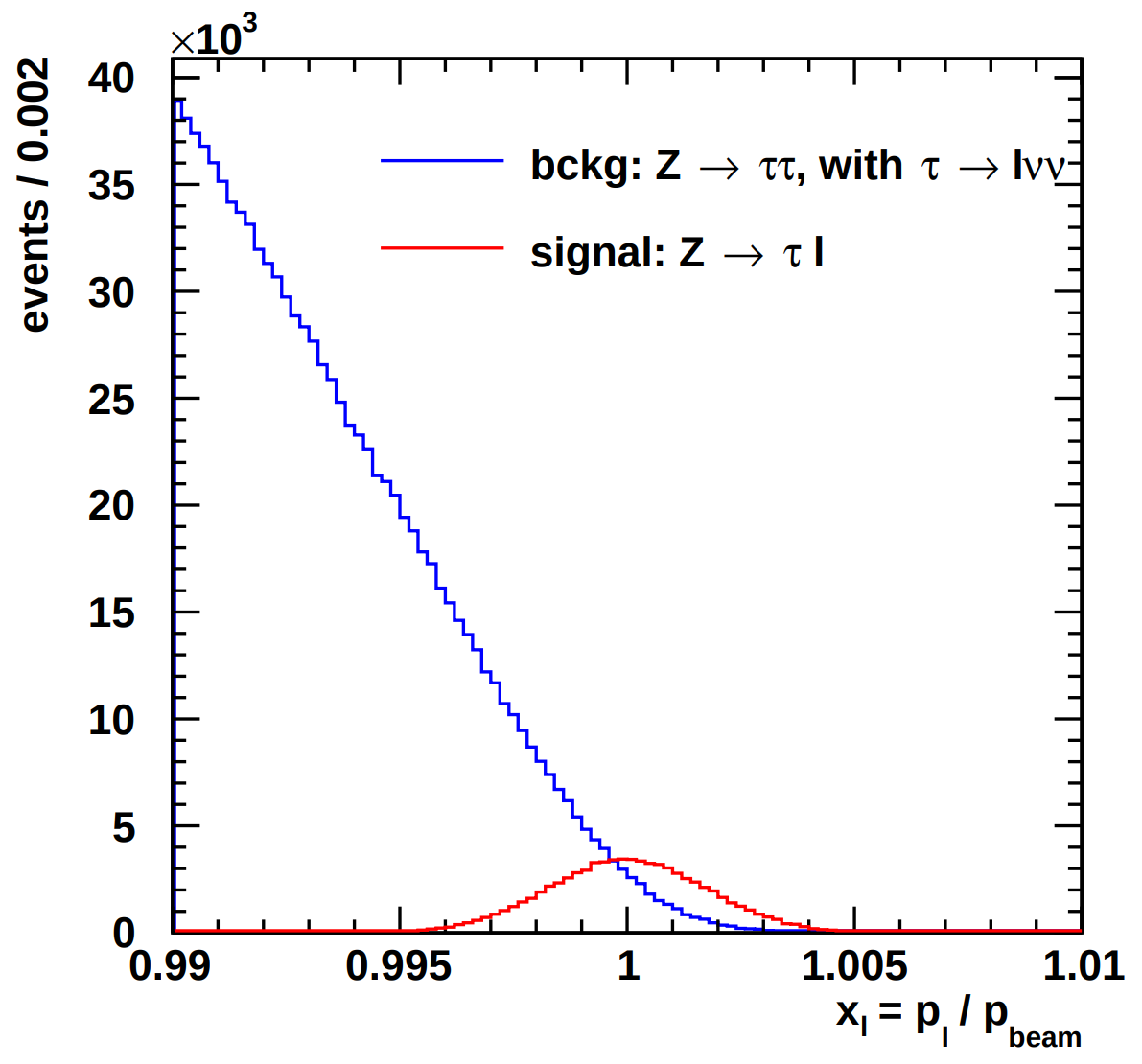}
\caption{Momentum distribution of the final state lepton in the FCC-ee search for the lepton flavor violating decay $Z \to\tau l, l = e, \mu$. Taken from Ref. \cite{Dam:2018rfz}.} \label{Ztaul}
\end{figure}

The cLFV can is studied in decays of the Z boson. 
The improvement compared to the HL-LHC is expected to be significantly better for the LFV Z boson decays for the branching fractions of $Z \to e\mu$, $Z \to e\tau$, and $Z \to \mu\tau$.
The improvement depends on how well background from $Z \to\mu\mu$ decays with a muon mis-reconstructed as an electron can be controlled (e.g., using $dE/dx$ information). 
The expected momentum distribution of the final state lepton for the signal and background at FCC-ee conditions is shown in Figure~\ref{Ztaul}.
It may be possible to achieve up to three orders of magnitude improvement at the FCC-ee.

The FCC-ee will allow sensitive measurements of cLFV in $\tau$ decays.
The cLFV in $\tau$ decays can be enhanced by several orders of magnitude and, due to the heavy mass of $\tau$, more cLFV processes are kinematically allowed compared to the case with muons.
Particularly interesting are the decays of the $\tau \to 3\mu$ and $\tau\to\mu\gamma$.
With the excellent FCC-ee invariant mass resolution, the search for the $\tau \to 3\mu$ mode is expected to be essentially background free, and sensitivity down to a branching fraction of $\mathcal{O}(10^{-10})$ should be within reach.
The sensitivity to the $\tau\to \mu\gamma$ branching fraction at the FCC-ee is expected to reach about $2\times 10^{-9}$.

Several other cLFV tau decays can be studied at the FCC-ee, similarly to what has been achieved by the B factories. 
Particle identification, which will be available in the FCC-ee detectors, should make these measurements highly competitive with the ultimate precision achievable in Belle~II.

\begin{figure}[!th]
\centering
\includegraphics[width=\linewidth]{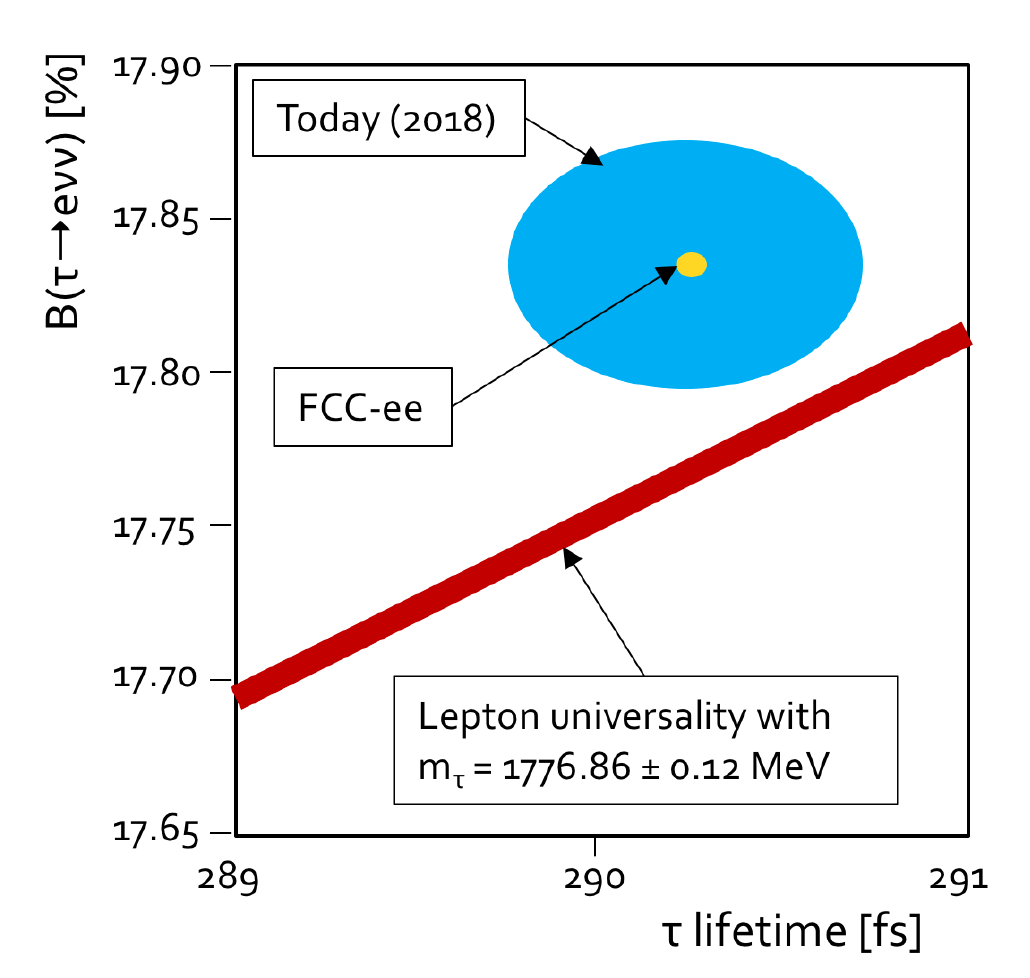}
\caption{Branching fraction of $\tau\to e\bar{\nu}_e\nu_\tau$ vs. $\tau$ lepton lifetime. The current world averages of the direct measurements are indicated with the blue ellipse, while the projected FCC-ee accuracy is provided with the small yellow ellipse. The red line corresponds to the prediction based on lepton universality given the present world-average value $\tau$ lepton mass, which may be further improved by the current and proposed charm factories.. Taken from \cite{Dam:2018rfz}.} \label{TauLifetime}
\end{figure}

Finally, the large $\tau$ samples expected at the FCC-ee should allow measurement of the $\tau$ lepton lifetime to an absolute precision of \SI{0.04}{\fs} and its leptonic branching fractions to an absolute precision of $3 \times 10^{-5}$ \cite{Dam:2018rfz}.
This would allow measurement of the Fermi constant in $\tau$ decays to a similar or even higher precision.
Comparing this number with the canonical $G_F$ measurement based on the muon lifetime offers another way of probing for new physics possibly responsible for non-flavor-universal couplings, as is shown in Figure~\ref{TauLifetime}.

Another direct test of lepton flavor universality can be achieved at the FCC-ee by the precision
simultaneous measurements of the branching fractions of the $\tau\to \mu\bar{\nu}_\mu\nu_\tau$ and $\tau\to e\bar{\nu}_e\nu_\tau$ decays.
The FCC-ee will be able to achieve the statistical precision of $\sim 10^{-5}$ in these branching fraction measurements.
These studies will be continued in the future, and they have the potential to provide the most stringent lepton universality tests achievable.

\section{Summary}
The FCC accelerator can open up wide-ranging physics program, allowing unprecedented tests of the Standard Model and further improving our knowledge of the fundamental interactions.
Among all proposed future lepton colliders, FCC-ee offers the largest sample for heavy-flavor studies. 
The FCC-ee program starts with collisions with center-of-mass energy at the Z boson mass with the highest possible luminosity, allowing collection of $5\times 10^{12}$ Z bosons in a few years.
This period will be a golden time for the precise measurements of weak interaction parameters and finding New Physics beyond the Standard Model of particle physics in rare decays.
Current results from flavor experiments may indicate the first hints of new physics, and 
FCC-ee Z-pole running will increase samples well beyond currently running experiments.
Extremely large data samples also allow important tests of lepton universality and searches for lepton flavor violation where $1-3$ orders of magnitude improvements are expected.
The outstanding physics opportunities make the heavy-flavor program one of the most exciting and attractive prospects at the FCC-ee.

\begin{acknowledgments}
I acknowledge support from the U.S. Department of Energy grant DE-SC0020255 and the National Science Foundation grant number 1906674. 

\end{acknowledgments}

\bigskip 

\end{document}